\documentclass[aps,prx,reprint,superscriptaddress,amsmath,amssymb]{revtex4-2}

\usepackage{graphicx}
\usepackage{dcolumn}
\usepackage{bm}
\usepackage{graphicx}       

\begin{document}

\title{{Relativistic Limits of Decoding: Critical Divergence of Kullback–Leibler Information and Free Energy}
}

\author{Tatsuaki Tsuruyama}
\email{tsuruyam@kuhp.kyoto-u.ac.jp}
\affiliation{Department of Physics, Tohoku University, Sendai 980-8578, Japan}
\affiliation{Department of Drug Discovery Medicine, Kyoto University, Kyoto 606-8501, Japan}
\affiliation{Department of Clinical Laboratory, Kyoto Tachibana University, Kyoto 607-8175, Japan}

\date{\today}

\begin{abstract}
We present a statistical-mechanical framework based on the Kullback–Leibler divergence (KLD) to analyze the relativistic limits of decoding time-encoded information from a moving source. By modeling the symbol durations as entropy-maximizing sequences and treating the decoding process as context-sensitive inference, we identify KLD between the sender and receiver distributions as a key indicator of contextual mismatch. We show that, under Lorentz transformations, this divergence grows with the sender's velocity and exhibits critical divergence as the velocity approaches the speed of light. Furthermore, we derive an analytic expression for the Fisher information and demonstrate that decoding sensitivity diverges similarly, indicating instability near the relativistic limit. By introducing an information-theoretic free energy based on the decoding cost, we determine a critical velocity beyond which decoding becomes thermodynamically impossible. These results reveal a phase-transition-like behavior in relativistic information transfer and provide a unified interpretation of KLD, Fisher information, and free energy as measures of decodability. The formalism developed here offers new insights into high-speed communication, relativistic signal processing, and information geometry in non-inertial frames.

\end{abstract}

\maketitle

\section{Introduction}

Understanding the fundamental limitations of information transmission in cases where the sender moves at relatively high speeds—requiring a relativistic framework—sheds light on the essential challenges not only in high-speed communication systems but also in understanding how information depends on the structure of space-time~\cite{Peres2004, Unruh1976}.

One of the key challenges in this setting is the distortion of temporal structure caused by the Lorentz transformation. A receiver lacking precise knowledge of the sender’s speed may misinterpret the duration and order of symbols, reconstructing an incorrect probability distribution over the transmitted message. This results in a breakdown of the intended semantic and contextual structure of the information~\cite{Brillouin2013, CoverThomas}.

To quantify this breakdown, we propose using the Kullback–Leibler divergence (KLD) between the sender's and receiver's symbol distributions\cite{Shannon1948}. KLD measures the degree of contextual inconsistency resulting from speed-induced time dilation and serves as a statistical indicator of irreversible decoding distortion. Although KLD has been widely employed in information geometry and Bayesian inference, its critical behavior under relativistic transformations has not been systematically investigated in the context of time-encoded communication.

We demonstrate that KLD diverges as the sender’s velocity approaches the speed of light, exhibiting a critical regime in which even small errors in velocity estimation can lead to catastrophic information loss. This divergence is accompanied by an explosion of the Fisher information, indicating instability in the decoder’s sensitivity to motion.

Furthermore, we argue that the receiver’s ability to decode is governed by the thermodynamic cost associated with restoring the sender’s context, and we introduce an effective free energy formalism. Within this framework, we derive a critical velocity at which the receiver’s free energy vanishes, beyond which decoding becomes thermodynamically impossible.

Our approach reveals that relativistic communication is constrained not only by velocity and channel capacity but also by geometric divergence in informational structure. The formalism presented here provides a unified perspective that links KLD, Fisher information, and information-theoretic free energy, offering new insights into relativistic signal processing and the limits of semantic recoverability in space-time\cite{ Amari2000}.

\section{Results}

\subsection{Statistical Model of Time-Encoded Information Sequences}

We consider a time-ordered sequence of information-bearing symbols
\[
\mathcal{X} = \{ x_k \}_{k=1}^{N},
\]
where each symbol \( x_k \) belongs to a finite alphabet
\[
\mathcal{A} = \{{\mathcal{A} }_j \}_{j=1}^{n},
\]
with each symbol type \( a_j \) associated with a duration (code length) \( \tau_j \) and an occurrence probability \( p_j \). The total number of transmitted symbols (sequence length) is \( N = |\mathcal{X}| \).

The average transmission time per symbol is defined as
\begin{equation}
\langle \tau \rangle := \sum_j p_j \tau_j = \mathrm{const.},
\end{equation}
and we consider entropy maximization under the constraint that this average transmission time takes a specified expected value. As will be shown later, this condition is equivalent to imposing that the total energy required for communication is held constant. Assuming that each symbol is transmitted as an optical pulse with constant average power \( P \), the energy required to transmit symbol type \( a_j \) is
\begin{equation}
\varepsilon_j = P \cdot \tau_j.
\end{equation}
The total energy required to transmit the sequence is then
\begin{equation}
E = \sum_{j=1}^{n} p_j \varepsilon_j = P \langle \tau \rangle = \mathrm{const.}
\end{equation}

The number of occurrences of each symbol type \( a_j \) within the sequence is
\[
N_j = N p_j,
\]
satisfying the normalization condition
\begin{equation}
\sum_{j=1}^{n} p_j = 1.
\end{equation}

The total number of distinguishable sequences with fixed counts \( \{ N_j \} \) is given by the multinomial coefficient
\begin{equation}
\Omega = \frac{N!}{\prod_{j=1}^{n} N_j!},
\end{equation}
and the corresponding (Shannon) entropy is
\begin{equation}
S := \log \Omega = -N \sum_{j=1}^{n} p_j \log p_j.
\end{equation}

To find the symbol probability distribution \( \{ p_j \} \) that maximizes the entropy under the constraints of normalization and a fixed average transmission time, we introduce Lagrange multipliers \( \alpha \) and \( \beta \) and consider the objective function:
\begin{equation}
\mathcal{L} = S + \alpha \left( \sum_{j=1}^{n} p_j - 1 \right) + \beta \left( \sum_{j=1}^{n} p_j \tau_j - \langle \tau \rangle \right).
\end{equation}

Solving the extremization condition (see Appendix~A), we obtain:
\begin{equation}
p_j = \frac{\exp(-\beta \tau_j)}{Z}, 
\quad
Z = \sum_{j=1}^{n} \exp(-\beta \tau_j),
\end{equation}
confirming that the maximum entropy distribution under a fixed average code duration constraint is exponential.

Using the exponential distribution and the constraint, the maximum entropy can be expressed as:
\begin{equation}
S_{\mathrm{max}} = -N \sum_{j=1}^{n} p_j \log p_j = N \beta \langle \tau \rangle,
\end{equation}
leading to:
\begin{equation}
\beta = \frac{S_{\mathrm{max}}}{N \langle \tau \rangle} = \mathrm{const.}
\end{equation}

Finally, the energy cost can be related to the entropy rate via:
\begin{equation}
E = P \langle \tau \rangle = \frac{P}{\beta} \cdot \beta \langle \tau \rangle = T_{\mathrm{info}} \cdot S_{\mathrm{max}},
\end{equation}
where we define the \textit{information thermodynamic temperature} as:
\begin{equation}
T_{\mathrm{info}} := \frac{P}{\beta}.
\end{equation}
Therefpre, we can interpret \(\beta\) as the inverse of the information thermodynamic temperature, representing the cost associated with transmitting longer-duration symbols under the entropy maximization constraint. This quantity will play a central role in analyzing the thermodynamic constraints of decoding in relativistic and finite-resource scenarios.

\subsection{Lorentz-Induced Transformation of Code Durations and Perceived Distributions}

We now consider the relativistic case in which the sender (system A) moves at a constant velocity $v$ with respect to the receiver (system B), who is at rest in an inertial frame. Each transmitted symbol $x_j$ has proper duration $\tau_j$ in the sender’s rest frame. Due to relativistic time dilation, the receiver observes each symbol with a contracted duration:
\begin{equation}
\tau_j^{(b)} = \frac{\tau_j^{(a)}}{\gamma(v)} = \frac{\tau_j}{\gamma(v)}, \quad \gamma(v) = \left(1 - \frac{v^2}{c^2} \right)^{-1/2}, \label{eq:lorentz_tau}
\end{equation}
where $c$ is the speed of light, and $\gamma(v)$ is the Lorentz factor.

Now suppose the receiver misestimates the sender’s velocity as $v_0$. Then, the receiver uses a Lorentz factor $\gamma_0 = \gamma(v_0)$ and interprets the transmitted code as
\begin{equation}
\tau_j^{(b)} = \frac{\gamma(v)}{\gamma_0} \tau_j. \label{eq:tau_misestimation}
\end{equation}
This results in a mismatch between the sender's intended symbol durations and the receiver's interpreted durations.

The sender’s symbol distribution is given by the exponential form
\begin{equation}
p_j^{(a)} = \frac{e^{-\beta \tau_j}}{Z^{(a)}}, \quad Z^{(a)} = \sum_{j=1}^{n} e^{-\beta \tau_j}. \label{eq:sender_dist}
\end{equation}

The receiver, who interprets the durations as $\tau_j^{(b)} = \gamma(v)/\gamma_0 \cdot \tau_j$, reconstructs the symbol distribution as
\begin{equation}
p_j^{(b)} = \frac{e^{-\beta \tau_j^{(b)}}}{Z^{(b)}} = \frac{e^{-\beta \frac{\gamma(v)}{\gamma_0} \tau_j}}{Z^{(b)}}, \label{eq:receiver_dist}
\end{equation}
with normalization constant
\begin{equation}
Z^{(b)} = \sum_{j=1}^{n} e^{-\beta \tau_j^{(b)}} = \sum_{j=1}^{n} e^{-\beta \frac{\gamma(v)}{\gamma_0} \tau_j}. \label{eq:Zb}
\end{equation}

Under the assumption that the receiver is unaware of the motion (i.e., $v_0 = 0$ and thus $\gamma_0 = 1$), the interpreted duration becomes
\begin{equation}
\tau_j^{(b)} = \gamma(v) \tau_j, \quad \Rightarrow \quad p_j^{(b)} = \frac{e^{-\beta \gamma(v) \tau_j}}{Z^{(b)}}. \label{eq:passive_receiver}
\end{equation}

The misalignment between $\tau_j$ and $\tau_j^{(b)}$ leads to a fundamental distortion in the code structure, where the same sequence of durations $\{\tau_j\}$ gives rise to different probability weights $\{p_j^{(a)}\}$ and $\{p_j^{(b)}\}$ in the two frames. This discrepancy is not merely observational but impacts the statistical geometry of the code space, as explored in the following sections.

To quantify the severity of this mismatch and its implications on information recoverability, we now introduce divergence measures between these two distributions and examine their behavior as functions of the sender’s velocity.

\subsection{Kullback--Leibler Divergence and Consistency Conditions}

To quantify the mismatch caused by relativistic time distortion, we introduce the Kullback--Leibler divergence (KLD) between the sender’s and receiver’s symbol distributions. Using the distributions defined in the previous section, the KLD from \( \{p_j^{(a)}\} \) to \( \{p_j^{(b)}\} \) is
\begin{equation}
D_{\mathrm{KL}}(v) = \sum_{j=1}^{n} p_j^{(b)} \log \left( \frac{p_j^{(b)}}{p_j^{(a)}} \right).
\end{equation}
This simplifies to
\begin{equation}
D_{\mathrm{KL}}(v) = \beta (1 - \gamma(v)) \langle \tau \rangle_b + \log \left( \frac{Z^{(a)}}{Z^{(b)}} \right),
\end{equation}
where \( \langle \tau \rangle_b = \sum_j p_j^{(b)} \tau_j \) and \( \gamma(v) = (1 - v^2/c^2)^{-1/2} \).

In the simplified case of uniformly scaled durations and absorbed normalization, this reduces to
\begin{equation}
D_{\mathrm{KL}}(v) = \left(1 - \frac{1}{\gamma(v)}\right) S^{(a)},
\end{equation}
with the sender’s entropy
\begin{equation}
S^{(a)} =\beta \langle \tau \rangle + \log Z^{(a)}.
\end{equation}

The plot is shown in Fig.~1.

\subsection{Fisher Information and Instability of Decoding Sensitivity}

To characterize the sensitivity of decoding performance to small errors in the estimation of the sender’s velocity $v$, we introduce the Fisher information with respect to $v$. It is defined as the second derivative of the Kullback--Leibler divergence with respect to velocity:
\begin{equation}
I(v) := \frac{d^2}{dv^2} D_{\mathrm{KL}}(v). \label{eq:fisher_def}
\end{equation}

 The first derivative is
\begin{align}
\frac{d}{dv} D_{\mathrm{KL}}(v)
&= S^{(a)} \cdot \frac{d}{dv} \left(1 - \frac{1}{\gamma(v)} \right)
= S^{(a)} \cdot \frac{1}{\gamma(v)^2} \cdot \frac{d\gamma(v)}{dv}. \label{eq:first_derivative}
\end{align}

Now recall that the Lorentz factor is
\begin{equation}
\gamma(v) = \left(1 - \frac{v^2}{c^2} \right)^{-1/2}, \label{eq:gamma_def}
\end{equation}
so its derivatives are
\begin{align}
\frac{d\gamma(v)}{dv} &= \frac{v}{c^2} \left(1 - \frac{v^2}{c^2} \right)^{-3/2}, \label{eq:gamma_prime} \\
\frac{d^2\gamma(v)}{dv^2} &= \frac{c^2 + 2v^2}{(c^2 - v^2)^2 \sqrt{1 - v^2/c^2}}. \label{eq:gamma_double_prime}
\end{align}

Substituting into the second derivative of $D_{\mathrm{KL}}(v)$, we obtain:
\begin{align}
I(v)
&= \frac{d^2}{dv^2} \left[ \left(1 - \frac{1}{\gamma(v)} \right) S^{(a)} \right] \nonumber \\
&= S^{(a)} \cdot \left[ \frac{2}{\gamma^3(v)} \left( \frac{d\gamma(v)}{dv} \right)^2 - \frac{1}{\gamma^2(v)} \cdot \frac{d^2 \gamma(v)}{dv^2} \right]. \label{eq:fisher_general}
\end{align}

Using Eqs.~(\ref{eq:gamma_prime}) and (\ref{eq:gamma_double_prime}) above, we can write this explicitly as:
\begin{equation}
I(v) = \beta\langle \tau \rangle\cdot \frac{c^2 + 2v^2}{(c^2 - v^2)^2 \sqrt{1 - v^2 / c^2}}, \label{eq:fisher_final}
\end{equation}

where we substituted an equation (22) in the sharp peak limit.

\paragraph*{Low-velocity limit:}
For $v \ll c$, we can expand the denominator as $1 - v^2 / c^2 \approx 1$, so
\begin{equation}
I(v) \approx \frac{\beta\langle \tau \rangle}{c^2}. \label{eq:fisher_nonrelativistic}
\end{equation}
Thus, the Fisher information becomes constant and independent of $v$ in the non-relativistic regime.

\paragraph*{High-velocity limit:}
As $v \to c$, the denominator $(c^2 - v^2)^2 \sqrt{1 - v^2/c^2}$ tends to zero, so
\begin{equation}
I(v) \to \infty. \label{eq:fisher_relativistic}
\end{equation}
This indicates that the decoder's sensitivity to estimation errors becomes infinitely large near the relativistic limit, implying a critical instability in decoding (Fig. 2).

\paragraph*{Cramér--Rao lower bound:}
According to estimation theory~\cite{Cramer1946}, the variance of any unbiased estimator $\hat{v}$ for velocity is bounded below by the inverse of the Fisher information:
\begin{equation}
\mathrm{Var}(\hat{v}) \geq \frac{1}{I(v)}. \label{eq:cramer_rao}
\end{equation}

As $v \to c$, we find $\mathrm{Var}(\hat{v}) \to 0$, suggesting diverging accuracy requirement to maintain decoding stability. This confirms that ultra-relativistic encoding requires exponentially precise velocity synchronization to avoid catastrophic information loss.

\subsection{Free Energy Asymmetry and the Critical Velocity for Decoding Failure}

We now analyze the thermodynamic implications of context mismatch between sender and receiver in terms of free energy. We define the information-theoretic free energy $F$ as the difference between the average energy required for transmission and the product of information temperature and entropy.

\paragraph*{Sender frame:}  
In the sender's rest frame, each symbol transmission costs energy $\varepsilon_j = P \cdot \tau_j$ and the average energy per symbol is $\langle \varepsilon \rangle = P \langle \tau \rangle$. The free energy is defined as
\begin{equation}
F_A := \langle \varepsilon \rangle - T_{\text{info}} \cdot S^{(a)} = P \langle \tau \rangle - \frac{P}{\beta} \cdot S^{(a)}, \label{eq:free_energy_sender}
\end{equation}
where $T_{\text{info}} = P / \beta$ is the information thermodynamic temperature.

\paragraph*{Receiver frame:}  
The receiver, using a different distribution $p_j^{(b)}$ due to Lorentz-transformed durations, reconstructs the entropy using the cross-entropy:
\begin{equation}
H(p^{(a)}, p^{(b)}) := - \sum_j p_j^{(a)} \log p_j^{(b)} = S^{(a)} + D_{\mathrm{KL}}(v). \label{eq:cross_entropy}
\end{equation}
The free energy in the receiver’s frame is
\begin{equation}
F_B := \langle \varepsilon \rangle - T_{\text{info}} \cdot H(p^{(a)}, p^{(b)}). \label{eq:free_energy_receiver}
\end{equation}
The mismatch in entropy due to context distortion introduces a free energy gap:
\begin{equation}
\Delta F := F_A - F_B = T_{\text{info}} \cdot D_{\mathrm{KL}}(v). \label{eq:free_energy_gap}
\end{equation}
This equation reflects the thermodynamic cost of misinterpreting the contextual structure.

\paragraph*{Critical velocity:}  
The receiver can recover the original semantic structure only if $F_B > 0$. The condition $F_B = 0$ defines a critical transition point beyond which decoding becomes thermodynamically infeasible:
\begin{equation}
F_B = 0 \quad \Leftrightarrow \quad H(p^{(a)}, p^{(b)}) = \frac{\langle \varepsilon \rangle}{T_{\text{info}}} = \beta \langle \tau \rangle. \label{eq:critical_cond}
\end{equation}

Recalling Eq.~(\ref{eq:cross_entropy}) and using $S^{(a)} = \beta \langle \tau \rangle + \log Z^{(a)}$, the condition becomes
\begin{equation}
S^{(a)} + D_{\mathrm{KL}}(v_{\text{crit}}) = \beta \langle \tau \rangle. \label{eq:critical_condition_detail}
\end{equation}
Solving for $D_{\mathrm{KL}}(v_{\text{crit}})$:
\begin{equation}
D_{\mathrm{KL}}(v_{\text{crit}}) = \beta \langle \tau \rangle - S^{(a)} = -\log Z^{(a)}. \label{eq:critical_KLD}
\end{equation}

Now using the expression for $D_{\mathrm{KL}}(v)$ in Eq.~(\ref{eq:KLD_recall}), we find the critical condition:
\begin{equation}
\left(1 - \frac{1}{\gamma(v_{\text{crit}})} \right) S^{(a)} = -\log Z^{(a)}. \label{eq:critical_gamma_cond}
\end{equation}
Rewriting in terms of $\gamma(v_{\text{crit}})$:
\begin{equation}
\gamma(v_{\text{crit}}) = \frac{S^{(a)}}{S^{(a)} + \log Z^{(a)}} = \frac{\beta \langle \tau \rangle + \log Z^{(a)}}{\beta \langle \tau \rangle}. \label{eq:gamma_crit}
\end{equation}

Using the relation $\gamma = (1 - v^2 / c^2)^{-1/2}$, we solve for $v_{\text{crit}}$:
\begin{equation}
v_{\text{crit}} = c \cdot \sqrt{1 - \left( \frac{\beta \langle \tau \rangle}{\beta \langle \tau \rangle + \log Z^{(a)}} \right)^2 }. \label{eq:vcrit_exact}
\end{equation}

\paragraph*{Approximate expression:}  
In the limit where the codebook has $n$ symbols and $Z^{(a)} \approx n$ (uniform durations), we write
\begin{equation}
\log Z^{(a)} \approx \log n, \label{eq:logZ_approx}
\end{equation}
so Eq.~(\ref{eq:vcrit_exact}) becomes:
\begin{equation}
v_{\text{crit}} \approx c \cdot \sqrt{1 - \left(1 + \frac{\log n}{\beta \langle \tau \rangle} \right)^{-2}}. \label{eq:vcrit_final}
\end{equation}

This expression shows that the critical velocity decreases with increasing code complexity (larger $n$), indicating that complex symbolic codes are more vulnerable to relativistic decoding breakdown.

\paragraph*{Interpretation of the decoding regimes:}
\begin{itemize}
    \item $v < v_{\text{crit}}$: $F_B > 0$ — decoding is thermodynamically feasible.
    \item $v = v_{\text{crit}}$: $F_B = 0$ — the boundary of information recoverability.
    \item $v > v_{\text{crit}}$: $F_B < 0$ — irreversible loss of context; original semantics unrecoverable.
\end{itemize}

In the supercritical regime ($v > v_{\text{crit}}$), although the receiver may assign a new codebook to the distorted durations, the original code semantics is fundamentally lost. This transition is mathematically sharp and physically irreversible under the given resource constraint, analogous to a thermodynamic phase transition.

\begin{figure}[t]
    \centering
    \includegraphics[width=0.45\textwidth]{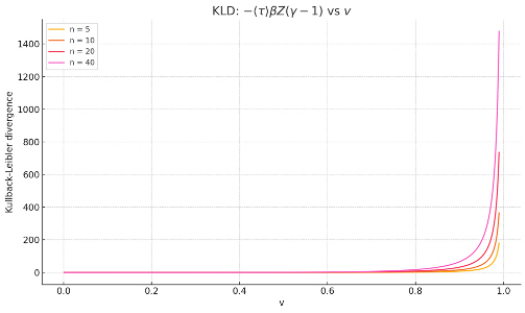}
    \caption{
    Kullback--Leibler divergence $D_{\mathrm{KL}}(v) = \left(1 - 1/\gamma(v)\right) S^{(a)}$ ~as a function of the sender’s velocity $v$, plotted for various codebook sizes $n = 5, 10, 20, 40$ with fixed parameters $\beta = 1.0$, $\langle \tau \rangle = 1.0$, and $c = 1$. As the sender's velocity increases, the Lorentz factor $\gamma(v)$ increases, resulting in a divergence in $D_{\mathrm{KL}}(v)$ near $v \to c$. This reflects the breakdown of contextual coherence between the sender and receiver frames. Larger $n$ leads to sharper divergence, indicating increased fragility of more complex codebooks under relativistic effects.
    }
    \label{fig:KLD}
\end{figure}

\begin{figure}[t]
    \centering
    \includegraphics[width=0.45\textwidth]{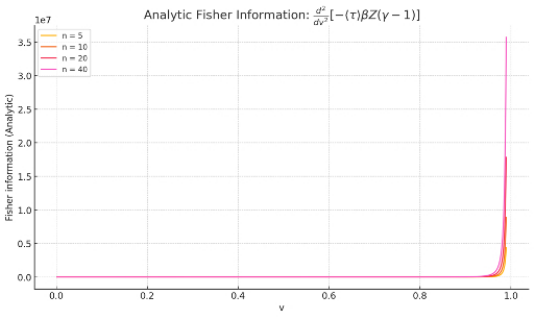}
    \caption{
    Fisher information $I(v) = \beta \langle \tau \rangle \cdot \frac{c^2 + 2v^2}{(c^2 - v^2)^2 \sqrt{1 - v^2/c^2}}$ plotted as a function of sender velocity $v$. As $v \to c$, $I(v)$ diverges, indicating that small errors in velocity estimation lead to drastically large distortions in the interpreted code structure. In the non-relativistic regime ($v \ll c$), Fisher information approaches the constant $\beta / c^2$. These results confirm that ultra-relativistic velocities induce critical instability in decoding.
    }
    \label{fig:Fisher}
\end{figure}

\begin{figure}[t]
    \centering
    \includegraphics[width=0.45\textwidth]{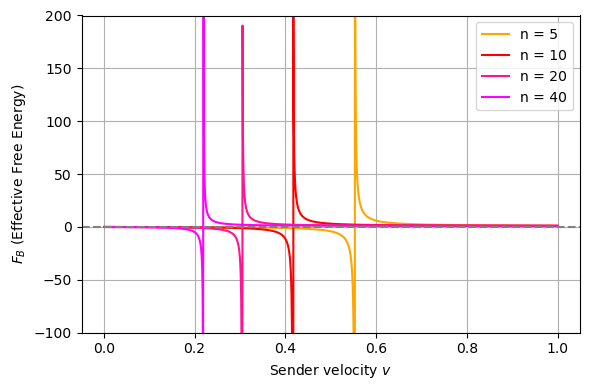}
    \caption{
    Receiver free energy $F_B(v)$ as a function of sender velocity $v$ for different codebook sizes $n = 5, 10, 20, 40$, under fixed $\beta = 1.0$, $\langle \tau \rangle = 1.0$, and $P = 1.0$. The free energy decreases with increasing $v$, reaching zero at the critical velocity $v_{\mathrm{crit}}$ (dotted vertical lines). This threshold defines the transition between thermodynamically reversible and irreversible decoding. As $n$ increases, $v_{\mathrm{crit}}$ decreases, reflecting the higher fragility of more complex symbolic codes under relativistic distortion.
    }
    \label{fig:FreeEnergy}
\end{figure}

\section{Discussion}

We addressed the fundamental issue of the space-time dependence of information by analyzing the decodability of time-encoded symbolic information under relative motion. By modeling the code sequence as an entropy-maximizing symbol distribution under a finite transmission time constraint, we demonstrated that the optimal distribution is exponential and is associated with an effective information temperature.

Our core result reveals that the relative motion between the sender and receiver induces a mismatch in the temporal context of symbols, which manifests as a distortion in the interpreted symbol distribution. This mismatch is quantified by the Kullback–Leibler divergence $D_{\mathrm{KL}}(v)$ between the sender's and receiver's code models. Crucially, we found that $D_{\mathrm{KL}}(v)$ diverges as the sender’s speed approaches the speed of light ($v \to c$), indicating a breakdown in contextual coherence.

To further characterize decoding sensitivity, we introduced the Fisher information $I(v)$ with respect to the sender's speed. We showed that $I(v)$ also diverges as $v \to c$, confirming that even minute errors in velocity estimation cause disproportionately large distortions in the receiver’s interpretation. This critical sensitivity signifies a phase-transition-like instability in the information geometry of relativistic codes.

From a thermodynamic perspective, we defined the free energy difference $\Delta F$ between the sender and receiver frames as a function of contextual mismatch. We derived a critical velocity $v_{\mathrm{crit}}$ beyond which the receiver's free energy vanishes ($F_B = 0$), marking a transition where decoding becomes thermodynamically infeasible given the available energy and entropy resources.

This critical behavior is reminiscent of a phase transition, in which the dynamics of a system change abruptly when a control parameter—here, the relative speed or code complexity—exceeds a threshold. Importantly, this analogy is not merely formal: we derived an explicit expression for $v_{\mathrm{crit}}$ and showed that it decreases inversely with the codebook complexity $n$, underscoring the vulnerability of rich symbolic structures to relativistic distortion.

These findings suggest that the Kullback–Leibler divergence between the sender and receiver models provides a unifying perspective on information geometry, statistical mechanics, and relativistic constraints, offering new insights into the fundamental limits of communication protocols in both high-speed and gravitationally curved spacetimes~\cite{Horowitz2014, Tsuruyama2025a}
.

\section{Conclusions}

We summarize the main contributions of this work as follows:

\begin{enumerate}
    \item We formulated a statistical model of time-encoded symbol sequences based on entropy maximization under a duration constraint, leading to exponential symbol distributions with an effective information temperature.

    \item We analyzed how Lorentz transformations distort the perceived symbol durations, resulting in a mismatch between sender and receiver distributions.

    \item This mismatch was quantified using the Kullback--Leibler divergence, which diverges as the sender's velocity approaches the speed of light, indicating critical instability in the decoding process.

    \item The Fisher information with respect to sender velocity was shown to diverge in the ultra-relativistic regime, revealing an extreme sensitivity of decoding accuracy to velocity estimation errors.

    \item We introduced a thermodynamic formulation of decoding based on the information free energy and derived a critical velocity $v_{\mathrm{crit}}$ at which decoding becomes energetically infeasible. The critical velocity was shown to decrease with increasing code complexity, demonstrating the fundamental limitations of high-speed symbolic communication.

\end{enumerate}

These results establish relativistic constraints on information decodability in a fully statistical and thermodynamic framework. They pave the way for future investigations of relativistic communication systems, quantum decoding limits, and the application of information geometry to curved spacetime environments, including near black hole horizons~\cite{Hawking1975, Tsuruyama2025b}

\section*{Acknowledgements}

This study was supported by a Grant-in-Aid from the Ministry of Education, Culture, Sports, Science, and Technology of Japan (Synergy of Fluctuation and Structure: Quest for Universal Laws in Nonequilibrium Systems, P2013-201).

\appendix

\section{Derivation of the Maximum Entropy Distribution}

We start with the entropy,
\begin{equation}
S = -N \sum_{j=1}^{n} p_j \log p_j,
\end{equation}
under the constraints:
\[
\sum_{j=1}^{n} p_j = 1, \quad T = N \sum_{j=1}^{n} p_j \tau_j,
\]
where \(N\) is the total number of symbols in the sequence.

We define the Lagrangian:
\begin{equation}
\mathcal{L} = -N \sum_{j=1}^{n} p_j \log p_j + \alpha \left( \sum_{j=1}^{n} p_j - 1 \right) + \beta \left( N \sum_{j=1}^{n} p_j \tau_j - T \right).
\end{equation}

Taking the derivative with respect to \(p_j\) and setting it to zero gives:
\begin{equation}
\frac{\partial \mathcal{L}}{\partial p_j} = -N(1 + \log p_j) + \alpha + \beta N \tau_j = 0,
\end{equation}
which leads to:
\begin{equation}
\log p_j = -1 + \frac{\alpha}{N} + \beta \tau_j,
\end{equation}
and hence,
\begin{equation}
p_j = K \cdot \exp(-\beta \tau_j),
\end{equation}
where
\begin{equation}
K = \exp \left( -1 + \frac{\alpha}{N} \right).
\end{equation}

The normalization condition determines \textit{K} as:
\begin{equation}
1 = \sum_{j=1}^{n} p_j = K \sum_{j=1}^{n} \exp(-\beta \tau_j),
\end{equation}
so
\begin{equation}
K= \frac{1}{Z},
\end{equation}
where we define the partition function,
\begin{equation}
Z := \sum_{j=1}^{n} \exp(-\beta \tau_j).
\end{equation}
Thus, the maximum entropy distribution is explicitly:
\begin{equation}
p_j = \frac{\exp(-\beta \tau_j)}{Z}.
\end{equation}
This confirms that under the fixed average code duration constraint, the distribution maximizing the entropy is exponential with respect to the code duration.

\section{Derivatives of the Lorentz Factor and Derivation of Fisher Information}

The Lorentz factor is defined as
\begin{equation}
\gamma(v) = \left(1 - \frac{v^2}{c^2} \right)^{-1/2}. \label{eq:gamma_def_appendix}
\end{equation}

We compute its first and second derivatives with respect to velocity $v$.

\subsection{First Derivative}

Differentiating Eq.~(\ref{eq:gamma_def_appendix}), we obtain:
\begin{align}
\frac{d \gamma(v)}{dv} 
&= \frac{1}{2} \left(1 - \frac{v^2}{c^2} \right)^{-3/2} \cdot \left( \frac{2v}{c^2} \right) \nonumber \\
&= \frac{v}{c^2} \left(1 - \frac{v^2}{c^2} \right)^{-3/2}. \label{eq:gamma_prime_appendix}
\end{align}

\subsection{Second Derivative}

Differentiating again:
\begin{align}
\frac{d^2 \gamma(v)}{dv^2} 
&= \frac{d}{dv} \left[ \frac{v}{c^2} \left(1 - \frac{v^2}{c^2} \right)^{-3/2} \right] \nonumber \\
&= \frac{1}{c^2} \left(1 - \frac{v^2}{c^2} \right)^{-3/2} 
+ \frac{v}{c^2} \cdot \frac{d}{dv} \left(1 - \frac{v^2}{c^2} \right)^{-3/2} \nonumber \\
&= \frac{1}{c^2} \left(1 - \frac{v^2}{c^2} \right)^{-3/2}
+ \frac{v}{c^2} \cdot \frac{3v}{c^2} \left(1 - \frac{v^2}{c^2} \right)^{-5/2} \nonumber \\
&= \left(1 - \frac{v^2}{c^2} \right)^{-5/2} \cdot \frac{1}{c^2} \left[ \left(1 - \frac{v^2}{c^2} \right) + \frac{3v^2}{c^2} \right] \nonumber \\
&= \frac{c^2 + 2v^2}{(c^2 - v^2)^2 \sqrt{1 - v^2 / c^2}}. \label{eq:gamma_double_prime_appendix}
\end{align}

\end{document}